\documentclass[aps,prl,preprint,superscriptaddress]{revtex4-1}
\usepackage{amsmath}    
\usepackage{graphicx}   
\usepackage{verbatim}   
\usepackage{color}      
\usepackage{hyperref}   
\usepackage{graphicx,color,psfrag}
\usepackage{bm}
\usepackage[all]{nowidow}
\begin{document}
\addtolength{\parskip}{\baselineskip}
\setlength{\parindent}{0in}
\title{From Stress Chains to Acoustic Emission }

\author{Ke Gao} 
\email{kegao@lanl.gov }
\affiliation{Geophysics, Los Alamos National Laboratory, Los Alamos, NM, USA }
\author{Robert Guyer}
\affiliation{Geophysics, Los Alamos National Laboratory, Los Alamos, NM, USA }
\affiliation{Physics, University of Nevada-Reno, NV, USA }
\author{Esteban Rougier}
\affiliation{Geophysics, Los Alamos National Laboratory, Los Alamos, NM, USA }
\author{Christopher X. Ren} 
\affiliation{Geophysics, Los Alamos National Laboratory, Los Alamos, NM, USA } 
\author{Paul A. Johnson}
\affiliation{Geophysics, Los Alamos National Laboratory, Los Alamos, NM, USA }

\date{\today}

\begin{abstract}
A numerical scheme using the combined finite-discrete element methods (FDEM) is employed to study a model of an earthquake system comprising a granular layer embedded in a formation.  When the formation is driven so as to shear the granular layer, a system of stress chains emerges.  The stress chains endow the layer with resistance to shear and on failure launch broadcasts into the formation.  These broadcasts, received as acoustic emission, provide a remote monitor of the state of the granular layer, of the earthquake system.
\end{abstract}

\pacs{123}

\maketitle
 
Low frequency earthquakes (LFEs) \cite{LFE}, non-volcanic tremor \cite{tremor,slow,SSC} and acoustic emission \cite{Scholz,ML2} are examples of weak seismic signals that may serve as harbingers of a major seismic event, an earthquake \cite{Obara253}.  In a laboratory setting \cite{PSU1}, in which stick-slip events are simulated, acoustic emissions are detected away from the stick-slip events \cite{PSU2}.  Recent machine learning and related studies \cite{ML1,Brzinski2018} of the acoustic emissions are able to use them to predict location in the seismic cycle, i.e., the evolution of the acoustic emission as a stick-slip scenario unfolds allows one to follow the scenario and anticipate the subsequent earthquake.    In all  cases, field and laboratory, the signals involved are sourced or detected in a volume remote from the volume that spawns the earthquake.  These findings bring to the fore the question of the causal relationship between signals detected on passive, remote monitors and the dynamics of the elastic structures that launch important seismic events.  In this paper we introduce a numerical model that lets us follow this causality, i.e., examine/connect the dynamics in a granular system (fault gouge), to signals detected on passive remote monitors.  We find that stress chains \cite{AT,KD} are the principals in the dynamics of the granular system and that their dynamics are the source of acoustic emissions.  The numerical model necessarily combines discrete element methods (DEM), to describe grain-grain interactions, and finite element 
methods (FEM), to describe elasticity within grains and wave propagation away from the granular system \cite{DEM,Omid,FDEM}.  

\begin{figure} \includegraphics[width =12.9cm]{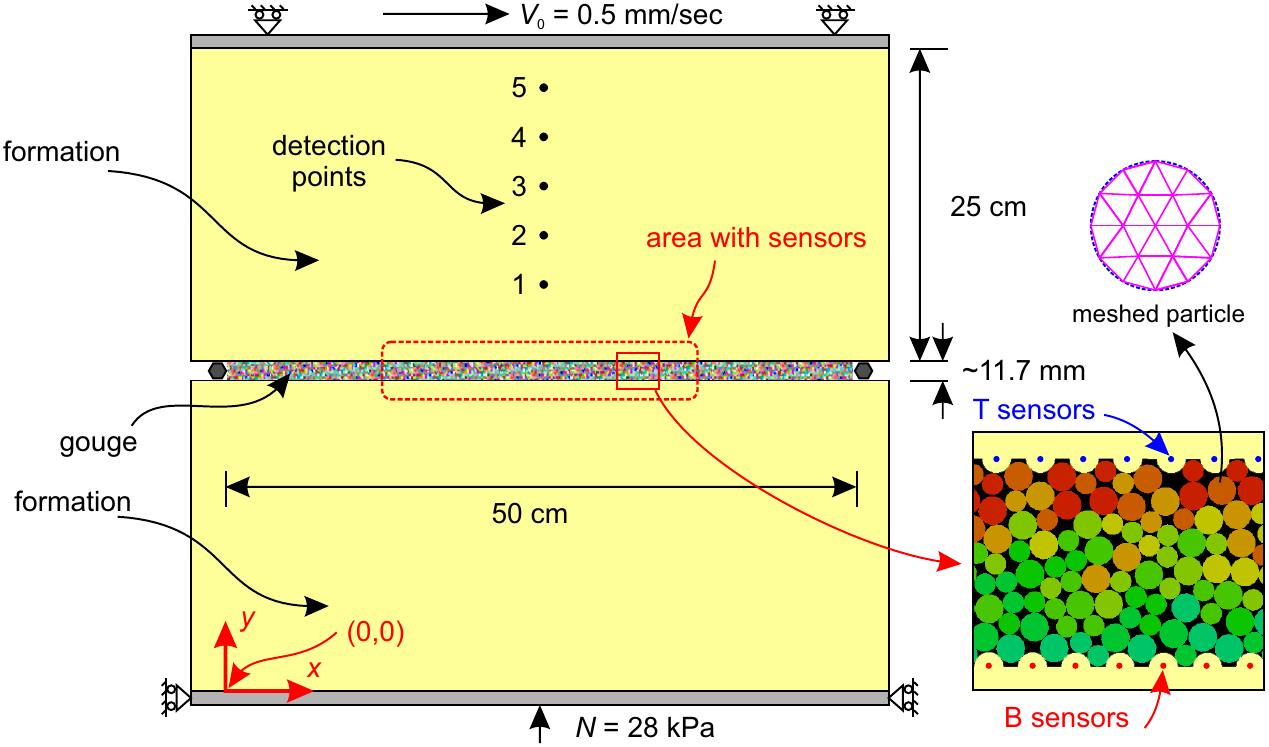} 
\caption{Model system.  The formation and the interior of the particles is described with a FEM scheme.  The contacts of gouge particles with one another and with the formation are described with a DEM scheme.  The formation has a rough surface profile (lower right).  Sensors in the formation surface, adjacent to the gouge, are spaced by $0.24$~mm.  The colors of particles represent the particle number.  The elastic properties of gouge and formation are listed in the Supplementary Material.} 
\label{system}
\end{figure}

The numerical model, two dimensional, comprises of a gouge layer of approximately $50\times 1$~cm$^2$ sandwiched between two $50\times 25$~cm$^2$ pieces of formation \cite{Ecke} (Fig.~\ref{system}).  The gouge layer is filled with particles (disks) of two diameters, $1.2$ and $1.6$~mm, and each particle is represented by 24 approximately equal size constant strain triangular FEM elements.  The particles are in contact with one another and with the formation via a penalty function based contact interaction algorithm \cite{munjiza2011computational}. The bottom edge of the lower formation piece is essentially rigid and fixed in $x$, and upon which a normal force, $N$, is applied.  The top edge of the upper formation piece, also essentially rigid and fixed in $y$, is driven uniformly at constant velocity $V_0$ in the $x$-direction.  The material density and elastic properties of gouge, formation, and top (bottom) edge of the formation piece are recorded in Table I of the Supplementary Material.  

\begin{figure} \includegraphics[width =8.8cm]{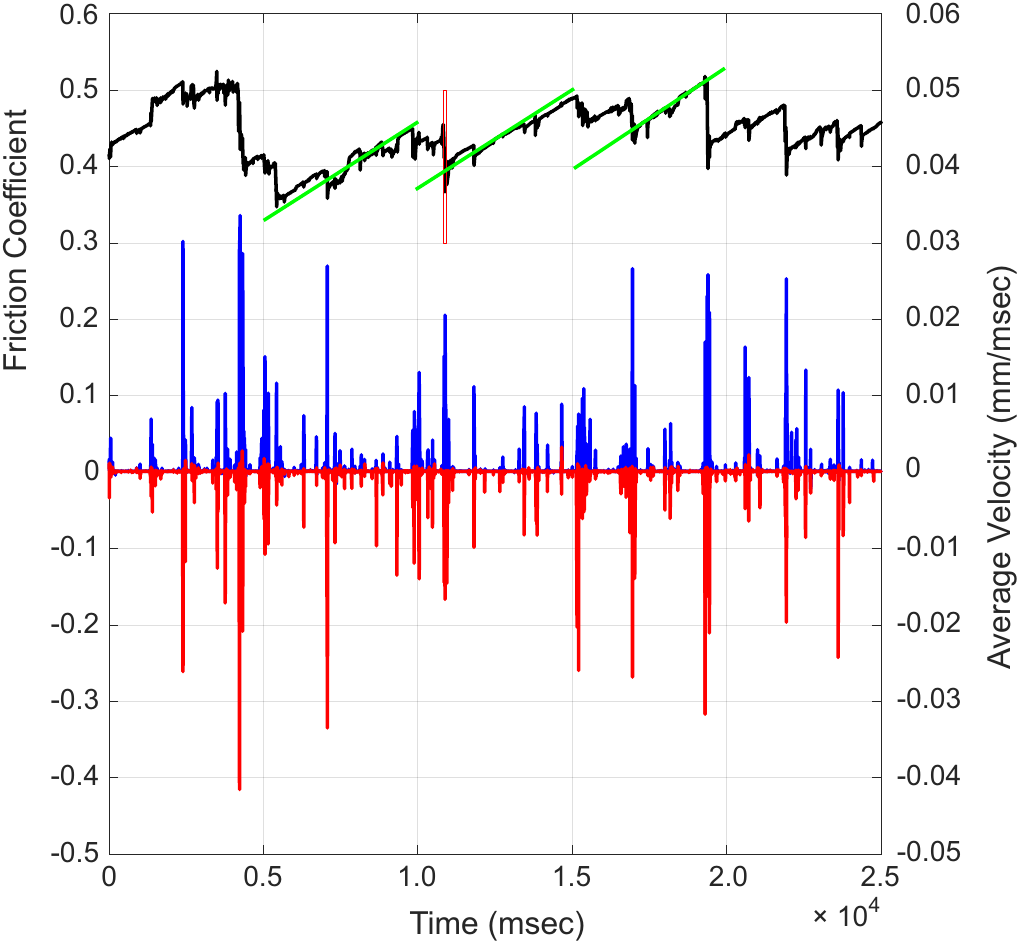} 
\caption{Primary outputs.  The coefficient of friction (black) as a function of time, left hand scale.  The average velocity of the top (T, blue) and bottom (B, red) sensors as a function of time, Eq.~(\ref{eq:Vbar}), right hand scale.  The green lines, all of the same slope, show the average elasticity of the gouge.  The stress drop near $11000$ msec (marked by red rectangle) is examined in Fig.~\ref{big}.} 
\label{basic}
\end{figure}

Typically the state of the system is set by a choice of $V_0$ and $N$: in the example here $V_0=0.50$~mm/sec and $N=28$~kPa.  The basic output is the shear and normal forces between the formation and gouge \cite{KG1}.  This is shown in Fig.~\ref{basic} as a coefficient of friction (black), $\mu=$ (shear force)$/$(normal force) vs time.  This stress-time pattern has several qualities of note.  There is on average a linear stress-time relationship that is punctuated by many small and a few large stress drops.  Compressive structures form in the gouge that push back against the effort of the upper formation piece to drag the lower formation piece along \cite{AT,KD}.  The three green lines, all with the same slope, illustrate the underlying spring-like character of the elasticity of the gouge.  Large stress drops occur irregularly in time.  The stress values at which large stress drops occur vary markedly, i.e., there is no single stress at which the system fails.  

\begin{figure} \includegraphics[width =12.9cm]{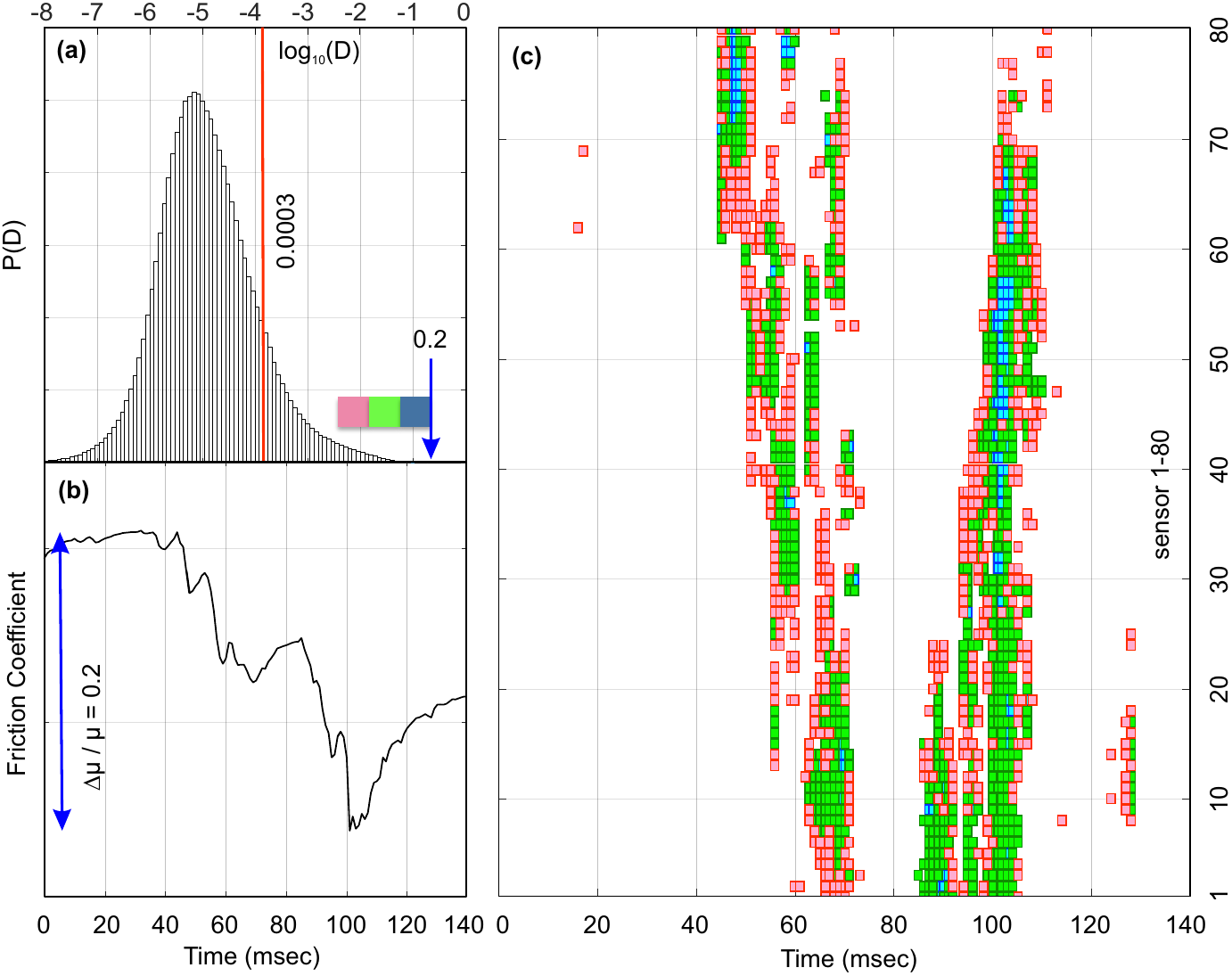} 
\caption{Spectrum of velocity dipole strengths, $D$. (a) The spectrum of velocity dipole strengths, for all space-time points, Eq.~(\ref{eq:D}), is broadly distributed from $10^{-8}$ to approximately $0.2$.  The space-time points $(n,t)$ with large $D$ (in the 3 intervals red, green, blue on the lower right in (a)) are used to identify events.  The stress near $11000$ msec in Fig.~\ref{basic} is shown in detail in (b).  The velocity dipole strength at $(n,t)$ for this time interval is shown in (c) where the time axis is the same as in (b) and $n$ is on the vertical axis.  The velocity dipole strength is shown with the color coding from (a).} 
\label{big}
\end{figure}

To begin to look at the dynamics of the system we look at the behavior of the formation immediately adjacent to the gouge.  At the top of the gouge (T) (at the bottom of the upper formation piece) and  at the bottom of the gouge (B) (at the top of the lower formation piece) there are 80 uniformly spaced sensor pairs that detect the motion of the formation, i.e., the velocity of the formation point with which the sensor is identified.  In Fig.~\ref{basic} we plot the average $x$-velocity of 80 T-sensors (sensors just above the gouge and near its center) as a function of time, $\overline{V}_X^T(t)$ (blue)
\begin{equation}
\overline{V}_X^T(t)=\frac{1}{80}\sum_{n=1}^{80}V_X^T(t,n),
\label{eq:Vbar}
\end{equation} 
where $V_X^T(t,n)$ is the $x$-component of the velocity of T sensor $n$ at time $t$.  The velocity  $\overline{V}_X^T(t)$ has a background value of approximately $0.25$~mm/sec ($V_0/2$) from which there are spikes in velocity that coincide with sharp, large ($1$~mm/msec) stress changes. In Fig.~\ref{basic}  we also plot the average $x$-velocity of the 80 associated B-sensors (sensors just below the gouge and near its center) as a function of time, $\overline{V}_X^B(t)$ (red).  The velocity $\overline{V}_X^B(t)$ also has a background velocity of approximately $0.25$~mm/sec.   From this background there are spikes in $\overline{V}_X^B(t)$ that coincide with sharp, large stress changes.  The spikes in $\overline{V}_X^B(t)$ are {\it opposite} in sign from the spikes in $\overline{V}_X^T(t)$.  That is, attending sharp, large stress changes are sharp, large  velocity dipoles delivered to the formation.  We focus on these velocity dipoles.  To characterize them we form the velocity dipole field
\begin{equation}
D(t,n)=\frac{V_X^T(t,n)-V_X^B(t,n)}{2}
\label{eq:D}
\end{equation}
that has a value at each moment of time $t$ at each sensor $n$, see the Supplementary Material.  The spectrum of values of $D(t,n)$, shown in Fig.~\ref{big}(a), is broadly distributed, $10^{-8}<D(t,n)<0.2$.  We use the magnitude of the velocity dipole field to form a detailed picture of the space-time structure of events in the gouge.  This is illustrated in Figs.~\ref{big}(b) and (c) for the large stress drop ($\Delta \mu/\mu\approx 0.2$) near $11000$ msec in Fig.~\ref{basic}.  Figure \ref{big}(b) is a zoom of the stress drop; Fig.~\ref{big}(c) shows the space-time points at which the velocity dipole field is large.  The sharp drop in Fig.~\ref{basic} is ragged in close up, Fig.~\ref{big}(b), and involves a complex set of events throughout the gouge,
Fig.~\ref{big}(c).  

\begin{figure} \includegraphics[width =12.9cm]{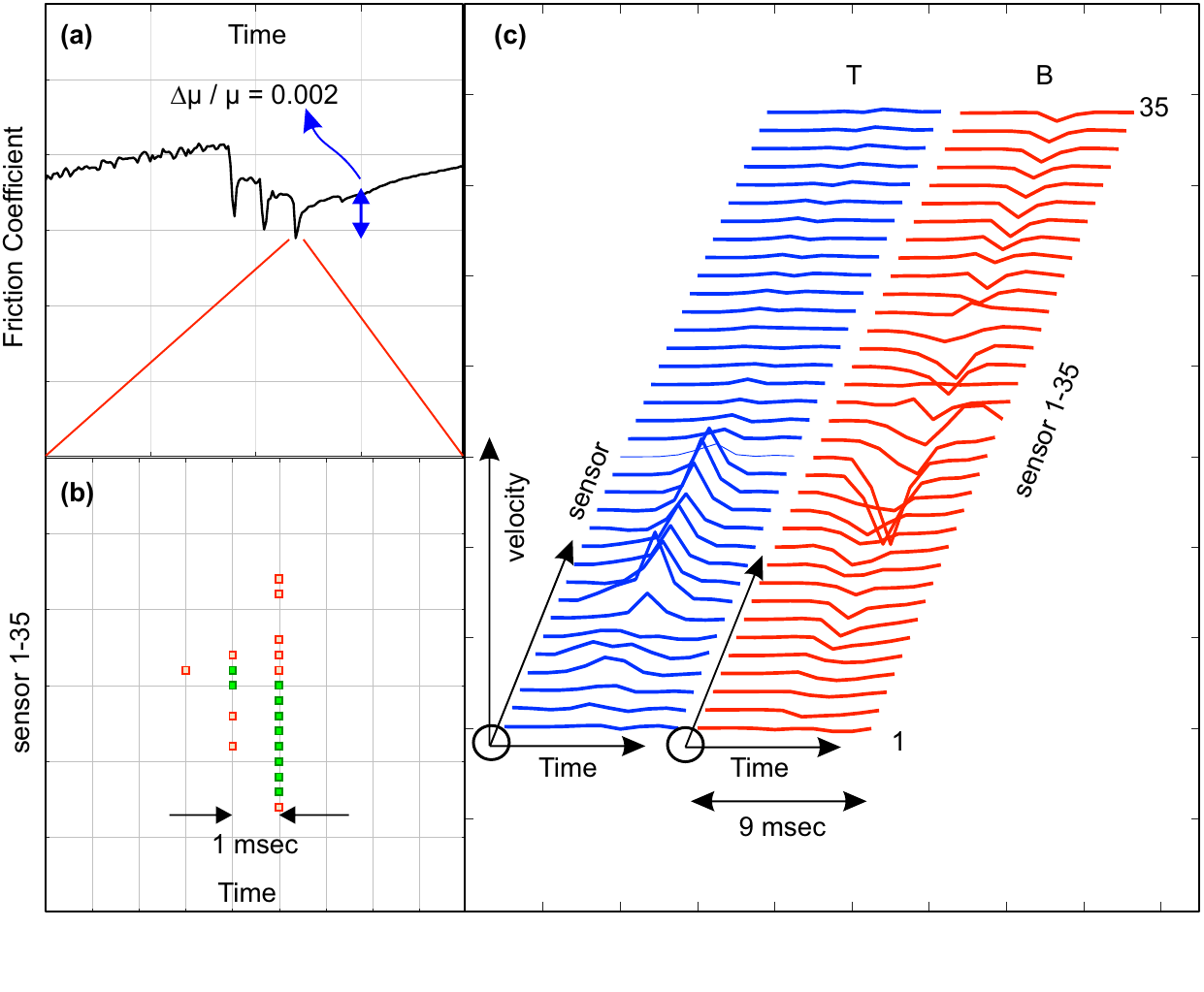} 
\caption{Details of a small, simple event.  (a) Friction coefficient vs time for 3 adjacent small events that have stress drop a few percent of a major event such as the one in Fig.~\ref{big}(b).  (b) The velocity dipole strengths associated with the third event in (a), color coded from Fig.~\ref{big}(a).  (c) Velocity dipole composition: the velocity of the top sensor set (blue) and, displaced to the right for clarity, and the velocity of the bottom sensor set (red).} 
\label{small}
\end{figure}

\begin{figure} \includegraphics[width =8.8cm]{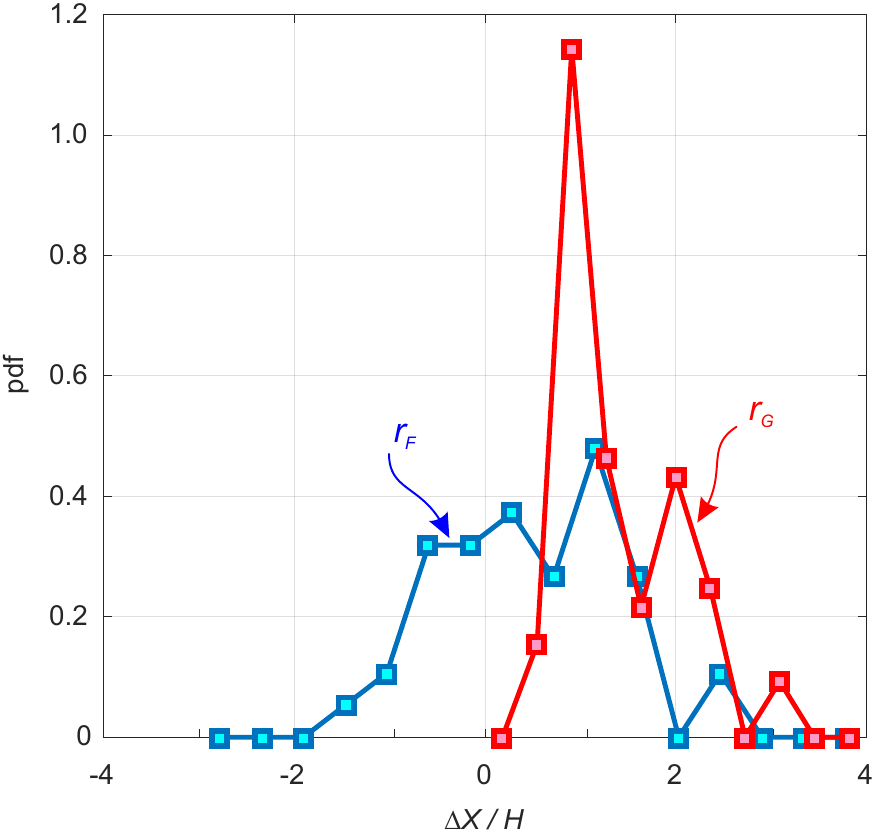} 
\caption{Probability distribution functions.  Probability distribution function of $r_F$ for the separation of the two components of the velocity dipole field (small events), blue.  Probability distribution function of $r_G$ for isolated stress chains, red.} 
\label{histcoup}
\end{figure}

We can also use the large values of $D(t,n)$ advantageously to locate and examine small, simple events.  In Fig.~\ref{small}(a) there are 3 stress drops of modest size ($\Delta\mu/\mu\approx 0.002$, about $1\%$ of that in Fig.~\ref{big}).  We look at the third stress drop in Fig.~\ref{small}(a).  This stress drop occurs rapidly in time, lasting approximately $3$~msec, and locally in space, involving large $D(t,n)$ at about 20 sensors, Fig.~\ref{small}(b).  To see more details we examine
$V_X^T(t,n)$ and $V_X^B(t,n)$ separately in Fig~\ref{small}(c).  We see that while $V_X^T(t,n)$ and $V_X^B(t,n)$ are simultaneous in time they are structured in space.  The B sensors contributing to the velocity dipole field are further along in $x$ than the associated T sensors.  The structural feature in the gouge, that upon failure delivers the forces that produce the velocity dipole field in the formation, acts like a strut.  It is tethered to the formation so that it can deliver a shear stress that crosses the gouge and gives the gouge (an unconsolidated material) a shear modulus.  We characterize the spatial structure of the velocity dipole field associated with small events like that in Fig.~\ref{small}(b) with a measure of the separation of the points of tether, $\Delta n=-n_X^B-n_X^T$,
\begin{equation}
n_X^{*}=\frac{\sum_n~n~d_X^{*}(n)}{\sum_n~d_X^{*}(n)},~~~d_X^{*}(n)=\int dt ~V_X^{*}(t,n),
\end{equation}
where the integral on $t$ and the sum on $n$ are over a domain that surrounds the small event, $d_X^{*}(n)$ is the displacement of sensor $n$, and $n_X^B$ ($n_X^T$) is typically negative (positive).  For the spectrum of $r_F=0.24\Delta n/H$ we find the result shown in Fig.~\ref{histcoup}, where $H$ is the nominal gouge layer thickness and $0.24$~mm is the spacing between sensors.  While the average value of $r_F$ is $\approx 0.5$, and there are both $+$ and $-$ values of $r_F$.  

\begin{figure} \includegraphics[width =8.8cm]{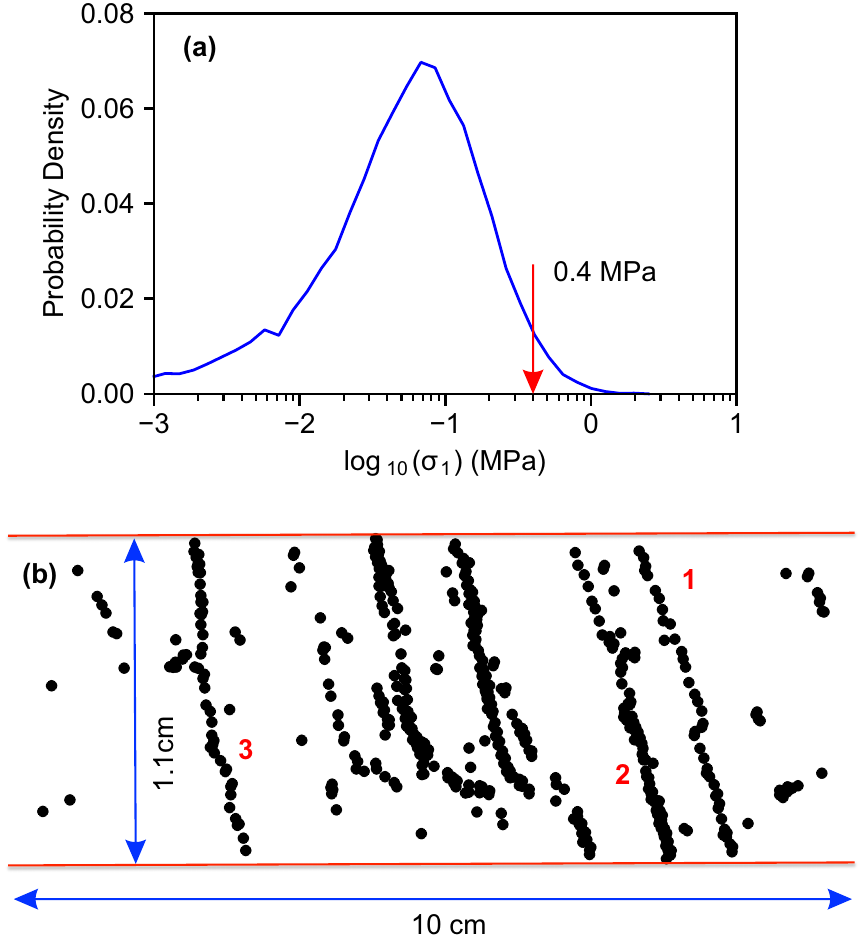} 
\caption{Stress chains.  (a) The spectrum of maximum principal stress of all elements in the gouge.  (b) Elements in the FEM description of the particles, with maximum principal stress greater than $0.4$~MPa, are shown in black according to their centroid positions.  Two chains of elements at the right, 1 and 2, are judged to be isolated as is the chain 3 on the left.  These chains are among the chains used to form the pdf of $r_G$ in Fig.~\ref{histcoup}. } 
\label{sc}
\end{figure}

We turn from examination of the motion of the formation, as evidenced in the behavior of the velocity dipole field, to the examination of forces in the gouge.  To do this, for each element in a gouge particle at each time $t$, we find the maximum principal stress.  We take this stress to characterize the force the element is carrying.  When the elements carrying large stress are illuminated an oriented fabric of forces crossing the gouge is revealed, Fig.~\ref{sc}.  Isolated stress chains are an important component of this fabric. These chains are oriented approximately as suggested by the velocity dipole field, e.g., Fig.~\ref{small}(c).  To make comparison of the force fabric with the velocity dipole field we characterize isolated stress chains that cross the gouge with $r_G=\Delta  X/H$, where $\Delta X$ is the projection of the chain figure onto the $x$-axis and  $H$ is  the width of the gouge. The spectrum of $r_G$ for isolated stress chains is shown in Fig.~\ref{histcoup}. 

The velocity dipole field has been used to examine events that are local in space and time. The two components of the velocity dipole field reveal detail about the structure in the gouge that produced the event.  Often forces are delivered to the T sensors that are upstream from the forces delivered to the B sensors.  This orientation mirrors the orientation of the fabric of forces in the gouge, Fig.~\ref{sc}.  The most extreme components of this fabric are the isolated linkages that reach from T to B.  From Fig.~\ref{histcoup}  and Fig.~\ref{sc} there are more complex structures in the force fabric.  On failure these structures deliver forces to T and B that are not easily identified with a simple geometry.  Comparison of the spectrum of $r_G$ and the spectrum of $r_F$ confirms that the geometry of the stress chains is closely related to the geometry of the velocity dipole field.

We complete the argument, stress chain $\rightarrow$ acoustic emission, by examining the motion of points in the formation far from the gouge, i.e., far from the velocity dipole field.  This is done in Fig.~S4 of the Supplementary Material, where it is seen that points in the formation remote from the gouge move much the same way as the velocity dipole field.  The causality is stress chain $\rightarrow$ velocity dipole field $\rightarrow$ far field signal, i.e., acoustic emission. The complexity of the numerical model, gouge plus formation, means that the simulations are quite long.  Consequently the formation does not have a true far field points.  The explicit near field demonstration in Fig.~S4 is complemented by use of the representation theorems in Aki and Richards \cite{AR} that would have the velocity dipole field as source for broadcast into the formation.

By the evidence of Fig.~\ref{sc}(b) one might estimate the stress chains to be separated by approximately the average of their projection onto the x-axis; a number of order $H$.  This suggests that on average about $50$ stress chains live in the gouge.  A single stress chain on failure releases stress that is a few percent of the stress drop associated with a large slip. This is consistent with a large slip involving almost all of the extant stress chains, comparing Fig.~\ref{big} and Fig.~\ref{small}.  In addition to assigning stress chains responsibility for time-sharp events we assign the elasticity evidenced in the average rise in stress seen in Fig.~\ref{basic} to their compression.

To conclude, we have employed a FDEM treatment of a numerical model of a sheared gouge layer.  The formation adjacent to the gouge and the interior of the particles that comprise the gouge have linear elasticity that is resolved with an FEM treatment.  Interactions of gouge particles with one another and with the formation, involving repulsive-only forces, are resolved with a DEM treatment.  When at fixed normal force the system is sheared at constant drive velocity the basic stress-time behavior shows intermittent large stress drops separated by intervals of approximately uniform elastic compression.  The principal actors in the gouge are stress chains.  Two sets of sensors at the gouge-formation interfaces allow determination of the velocity dipole field which describes the disturbance to the formation caused by the activity in the gouge.  The motion of the formation far from the gouge, taken to be the acoustic emission, is monitored.  We establish the linkage between stress chains, the velocity dipole field, and the acoustic emission.  That is, the relationship between the activity in the gouge and the far field.  This is an example of a causal scenario that one might hope to encounter and take advantage of in geophysical systems.  Non volcanic tremor, LFEs, etc. are far fields that may have a role in such a scenario.  The great complexity of geophysical elastic systems makes the demonstration of such a scenario very difficult.  The rapid increase in the quality and quantity of reliable geophysical monitoring and the infusion of new, sophisticated analysis methods offer promise that such scenarios are foreseeable. It is worth noting that acoustic emissions are coming dominantly from the force chain breakage but not necessarily entirely. Also, the system studied is representative of faults containing gouge; however, in real faults, acoustic emissions may also arise from block to block asperity slip and potentially other mechanisms such as material breakage as the slip front propagates.

\begin{acknowledgments}
{We acknowledge funding from Institutional Support (LDRD) at Los Alamos National Laboratory, as well as funding from the US DOE Office of Science, Geosciences. Technical support and computational resources from the Los Alamos National Laboratory Institutional Computing Program are highly appreciated.}
\end{acknowledgments}

\bibliography{SC2AEtemplate}

\newpage

\newpage

\end{document}